\documentstyle[12pt, aasms4, graphics]{article}

\begin{document}

\title{GAMMA RAY BURSTS HAVE MILLISECOND VARIABILITY}

\author{Katharine C. Walker \altaffilmark{1}, Bradley E. Schaefer
\altaffilmark{2} 
\altaffiltext{1}{katharine.walker@yale.edu}
\altaffiltext{2}{schaefer@grb2.physics.yale.edu}
}

\affil{Yale University, PO Box 208121, New Haven CT 06520-8121}

\author{and E. E. Fenimore\altaffilmark{3}
\altaffiltext{3}{efenimore@lanl.gov}
}

\affil{Los Alamos National Laboratory, MS D436, Los Alamos, NM 87545}

\begin{abstract}

We have performed searches for isolated flares and for steady flickering 
in Gamma Ray Burst light curves on the microsecond to millisecond time 
scales. Two bursts out of our sample of 20 revealed four isolated flares 
with time scales from $256 \mu s$ to $2048  \mu s$.  A wavelet analysis
for our 
sample showed low level flickering for all bursts on time scales from $256  
\mu s$ to 33 ms, with the majority of bursts containing rise times faster
than four milliseconds and 30\% having rise times faster than one
millisecond. 
These results show that millisecond variability is common in classical 
bursts and not some exceptional activity by a possibly separate class of 
bursts.  These fast rise times can be used to place severe limits on burst 
models:  (1) The characteristic thickness of the energy generation region
must be less than 1200 km along the line of sight.  (2) The angular size
of the gamma ray emission region 
as subtended from the central source must be less than 42 arc-seconds.
(3) 
The expanding ejecta must have a range of Lorentz factors along a
radius line with a dispersion 
of less than roughly 2\%.  (4) Within the external shock scenario, the 
characteristic dimension of the impacted cloud must be smaller than 16 
Astronomical Units on average.  (5) Within the collimated jet scenario, the 
collimation angle must be smaller than 42 arc-seconds.

\end{abstract}


\clearpage

\section{Introduction}

What is the shortest time scale of intensity variations in Gamma Ray Bursts 
(GRBs)?  This is an important question because this time scale can be used 
to place an upper limit on the size of the gamma ray emitting region.  
Historically, the rise time in the 5 March 1979 event was used to place a 
limit of $<300 km$ (Cline et al. 1980), although we now know that this
event 
was from a `galactic' Soft Gamma Repeater and hence irrelevant for 
cosmological GRBs.  Nevertheless, the basic argument remains in force for 
classical GRBs, with durations $<15 ms$ in the Konus catalog (Mazets et
al. 1981), and it provided one of the strong reasons to consider neutron
stars in burst models.

Since the launch of the Compton Gamma Ray Observatory, the Burst and Transient 
Source Experiment (BATSE) provides sufficient photons and time resolution to 
push a variability search to short time scales.  Bhat et al. (1992) 
demonstrated that GRB910711 has a total duration of $\sim 8 ms$, although
the claimed 0.2 ms spike detected in one BATSE detector is dubious since it is 
only 3-$\sigma$ in significance with many trials and since the spike is
not 
present in other BATSE detectors that should have seen it. Nevertheless, this 
event and others in the BATSE catalog with durations as short as 0.034 s 
(Fishman et al. 1994) show that some bursts have flares with durations as 
short as $\sim 8 ms$.  Mitrofanov (1989) suggested that bursts were
composed of microsecond flares such that dead time and pulse pile-up effects 
would greatly change burst demographics, but correlations between arrival 
times for photons in separate detectors shows that this possibility is not 
realized (Schaefer et al. 1992).  Similarly, with H. A. Leder, we have shown 
that photon energies are uncorrelated on microsecond time scales, so that 
burst flux can have only a small fraction of short duration blackbody 
emission.  Deng \& Schaefer (1997) did not find any coherent periodicities
from $16 \mu s$ and 33 ms in 20 of the brightest bursts.  Schaefer \&
Walker (1998) have discussed a spike in GRB920229 that has an e-folding
rise time of $220 \pm 130 \mu s$, a decay time of $400 \pm 100 \mu s$, a
significant spectral 
change over a time of $768 \mu s$, and a sharp spectral continuum feature
over a fractional energy range of 18\%.

The above results show that rare bursts can have light curve structure on 
time scales of $\sim 8 ms$ or even 0.22 ms.  But how exceptional are these
fast varying bursts? Are the rapid bursts a separate class whose limits 
cannot be applied to ordinary bursts? And what is the fastest time scale for 
ordinary bursts?  In this paper, we report on two separate searches for rapid 
variability in GRBs.  In the first search, we tested 20 bright 
bursts for the presence of isolated flares on time scales from $32-2048
\mu s$. In the second search, we use Haar wavelet transforms to evaluate the 
flickering activity in burst light curves on time scales from $2 \mu s$ to 
0.13 seconds.

\section{Isolated Flares}

One of the possible modes by which bursts can display rapid variability  
is to have isolated flares.  These might occur on any time scale and might 
be most prominent in either hard or soft photons.  Giles (1997) offers a 
reliable and efficient algorithm for searching a light curve for significant 
peaks.  In essence, his algorithm calculates a running mean and then seeks a 
significant deviation above this mean.  This algorithm searches through light 
curves which are successively binned by factors of two, so that 
we have tested light curves with bin sizes of $32 \mu s$, $64 \mu s$, $256
\mu s$, $512 \mu s$, $1024 \mu s$, and $2048 \mu s$. Our threshold is set
such that a flare 
would have to be more significant than $5 \sigma$ after accounting for all
the trials in a single burst.  We have modified this algorithm to reduce the 
size of the window used in the running average so as to minimize 
the effect of curvature in the overall shape of the light curve.

This isolated flare search was performed on BATSE TTE data, which is
perfect for rapid variability searches.  The TTE data records the arrival 
time (within a $2 \mu s$ time bin) and energy (within
four discriminator channels) of each photon.  The energy boundaries of
channels 1 
through 4 are roughly 25-50,  50-100, 100-300, and $>300$ keV. The on-board 
memory records only up to 32768 photons around the time of the BATSE trigger. 
 Typically, this quota of photons is used up in one or two seconds, which 
can only cover the leading portion of a long duration burst.  For short bursts, 
the entire episode might be in the TTE data, along with substantial times of 
only background light after the burst.  The time-tagged events are continuously
 written into a rotating memory so that TTE data is usually available for 
a fraction of a second before the BATSE triggers.  For times before the 
trigger, photons from all eight BATSE modules are recorded, although we have 
only used photons from triggered detectors.  The pulse pile-up time is 
$0.25 \mu s$ and the dead time is $0.13 \mu s$.

Our isolated flare search was performed on 20 of the brightest BATSE bursts 
(see Table 1).  These were chosen for the number of burst photons recorded  
in the short time interval during which TTE data is available.  Our set of 
bursts is a mixture of short intense bursts with fast variability completely 
covered by the TTE data and the brightest bursts of ordinary duration with 
high numbers of burst photons during the TTE data. The columns of Table 1 
gives the GRB name, the BATSE burst trigger number, the peak flux from 
50-300 keV over a 64 ms time bin, the T90 burst duration, the duration of 
the TTE data, and $<C[32]>$ the average count rate in $32 \mu s$ time
bins. We performed the tests on three separate light curve sets; with 
channels 1+2+3+4, channels 1+2, and channels 3+4.  Our search found only 
four significant flares in two bursts out of our sample of 20 bright GRB 
light curves.  

Our first burst with flares is the extremely bright GRB930131.  This
burst has an initial spike (with duration $\sim 1 s$) composed of two main
peaks (each with duration $\sim 0.1 s$) for which the first main peak has
two flares (of total durations $\sim 0.004$ and $\sim 0.01 s$) visible
only at the highest energies.  In channels 3+4, the light curve triggered 
on the $2048 \mu s$ time scale for each of the two flares on the first
main peak. The fast variations in this flare are primarily in channel 4, while 
channels 1 and 2 have no corresponding variations (see Figure 1c of 
Kouveliotou et al. 1994).  The spectrum of these flares are exceptionally hard.

Our second burst with flares is GRB920229.  This short burst has a 0.19 s 
duration, consisting of a smooth time-symmetric pulse followed by a spike 
with duration of roughly 0.003 s.  Within the spike, on the $256 \mu s$
time 
scale, our flare search triggered on a flare near the end (at our usual 
$5-\sigma$ threshold) as well as a flare near the beginning (although only at 
the $3-\sigma$ confidence level after allowing for all the trials
associated 
with our search for this one burst).  The e-folding rise time of this 
spike is $220 \pm 130 \mu s$, the e-folding fall time is $400 \pm 100 \mu
s$, while the spectrum significantly softens over a $768 \mu s$ time
interval during the 
spike's fall. The background subtracted count rate for the entire burst 
for channels 1, 2, 3, and 4 are 730, 1630, 2490, and 120 photons, which 
demonstrates a sharp spectral break around the energy boundary between 
channels 3 and 4.  Detailed spectroscopy shows the spectrum has a peak 
$\nu F_{\nu}$ at 200 keV with no significant flux above 239 keV, for a 
sharpness of $\Delta E/E = 18\%$.  These observations are presented in
detail in Schaefer \& Walker (1998).

This systematic study of 20 bright bursts shows that isolated flares of  
large amplitude are not common on the two millisecond time scale or faster.

\section{Flickering}

Another possible mode by which bursts can display rapid variability is to 
have many small amplitude flares flickering quietly.  This would just be an 
extension of the flickering seen on longer time scales as part of the multiple
 pulses forming the overall shape of many light curves.  What is the shortest 
time scale on which bursts flicker? Short duration flickers must fall below 
the thresholds already established by our isolated flare search, and this 
implies that the flickers are either isolated and of low relative amplitude 
or crowded together so that many flickers are bright at any one time.

If the low amplitude flares recur repeatedly, then there should be statistical 
evidence for the burst showing fluctuations above that expected from Poisson 
noise alone. One means to test for frequent low level fluctuations is a 
wavelet analysis. Wavelets are a set of mathematical functions that
form an orthonormal basis which can readily describe short duration events 
(Scargle 1997, Daubechies 1992).  Wavelets have already been used for analysis 
of GRBs on long time scales by Norris et al. (1994) and Kolaczyk (1997).

In particular, we have used the simple Haar wavelet, which is an antisymettric 
function consisting of one bin negative and the next bin positive with all  
other bins being zero.  For a given bin size, the wavelet activity is defined 
as the average of the squares of the product between the Haar wavelet and the 
light curve for all relative offsets.  To be quantitative, the Haar wavelet 
activity is equal to $<(C_{i}-C_{i+1})^{2}>$, where $C_{i}$ is the counts
in the `ith' time 
bin of the light curve and the angular brackets indicates an average over 
all values of i.  As such, the activity is a measure of the rise and fall  
times present in the light curve.  For normal Poisson variations alone, 
the expected activity level is $2<C_{i}>$. In practice, the observed
value is
slightly different due to dead time effects and the overall modulation of
the light 
curve on long time scales.  The RMS scatter of the Poisson activity is 
$(8/N)^{0.5}<C_{i}>$ where N is the number of time bins in the light
curve. Our normalized activity is the ratio between the observed activity
and that expected for Poisson variations alone.

The normalized activity is calculated for light curves with bin sizes  
varying by factors of two from $32 \mu s$ to 0.131 s.  In general, this
number 
is around unity for short bin sizes and it starts to rise significantly for 
some time scale which we identify as the shortest time scale of variability.  
From studies of simulated data, we find that the overall envelope of 
variability on long time scales does not produce activity on short 
time scales.  The existence of this shortest time scale of variability does 
 not imply either that all variations are on that time scale nor that the 
fast variations have high amplitude. Rather, there appears to be a 
continuum of variations ranging from large amplitude pulses of long duration 
to smaller pulses of short duration.

From studies of background data and of simulated data, we find that the  
normalized activity varies with a one-$\sigma$ scatter from 0.7\% to 3.0\%
for $<C[32]>$ values from 0.5 to 2.0.  This allows us to place a confident
limit on 
the shortest time scale of variability as the bin size in which the normalized 
activity is three-$\sigma$ above the Poisson level ($\tau_{min}$).  Such
time scales for 
each burst are tabulated in Table 2.
 
For GRB930131 and GRB920229, we recover the fast variations in the flares as 
$\tau_{min}$.  We find no significant correlation between $\tau_{min}$ and
either
$T_{90}$ or $<C[32]>$.  These facts indicate that the normalized activity
is indeed a 
measure of the shortest time scale of variability in a manner that is 
independent of brightness and duration.

What is the fractional amplitude of these flickers?  Let $A_{norm}$ be the
observed 
normalized activity, $V_{f}$ be the variance in the light curve due to
flickering, 
and $V_{p}$ be the variance in the light curve due to normal Poisson
fluctuations; 
then $A_{norm} = (V_{f}+V_{p})/V_{p}$. The variance of the flickering is
the square of the RMS amplitude for flickering in counts, $C_{f}$.  The 
variance from Poisson fluctuations equals the average number of counts in
the each bin of the light curve, $<C>$, which can be found by scaling from the 
$<C[32]>$ values in Table 2. Table 2 also lists the observed values of
$A_{norm}$ for the 33 ms light curve.  
The fractional amplitude of the flickers is then  $C_{f}/<C>
=([A_{norm}-1]/<C>)^{0.5}$. For the threshold time scales $\tau_{min}$, 
$A_{norm}\sim 1.06$ and $<C> \sim 100$ counts, so the flickers are
$\sim 2\%$ in amplitude.

Figure 1 displays the normalized activities as a function of the bin size for 
five bursts.  In these bursts and in all our 20 bright bursts, the
normalized 
activity is around unity for time scales less than $\tau_{min}$, and then
rises sharply above $\tau_{min}$.  

In some cases, the activity does not rise monotonically with time scale, for 
example the peak at 0.016 s for GRB930905 in Figure 1.  The time scale of 
these local maxima in normalized activity is $T_{peak}$, as tabulated in
Table 2. From our sample, we 
find significant peaks for seven bursts, with $T_{peak}$ 
ranging from 8.2 ms up to our highest observable value of 66 ms.  While it is 
possible that these peaks arise from flickers that have a characteristic rise 
time, we believe that the peaks are caused by single flares of large 
amplitude which contribute much activity on the time scale of their 
rise time. Indeed, with one exception (GRB930506, for which the peak has a 
small $A_{norm} = 1.24$), the $T_{peak}$ values can be linked to a single
specific rise with the same time scale.

While our isolated flare search measured durations, our wavelet activity 
search measured rise and fall times.  For several reasons we believe that our 
$\tau_{min}$ values are essentially rise times.  First, the $T_{peak}$
values for six 
bursts have been identified with particular rises.  Second, the
$\tau_{min}$ values 
for GRB920229 and GRB930131 correspond to specific rises in the light curve 
which are $> 2$ times faster than any significant fall. Third, in general,
bursts 
always display a substantially faster rise than fall (Barat et al. 
1984, Nemiroff et al. 1994).

Out of our 20 bright bursts, the range of $\tau_{min}$ is from $256 \mu s$
to 33 ms, with a median value of four milliseconds and 30\% with activity
at one millisecond or faster.  So we conclude that most burst light curves
contain rises with a time scales of order a millisecond.

\section{Implications} 

We have shown that the majority of GRBs have flickering with rise times faster 
than four milliseconds, while individual flares can vary with rise times as 
fast as $220 \mu s$.  Thus, millisecond variability is common in bursts,
and not just a rare phenomenon restricted to some special and possibly 
distinct class.

The rise and fall times measured by the wavelet activity can be used to  
place limits on GRB models.  Based on the recent discoveries of low energy 
counterparts (Costa et al. 1997, van Paradijs et al. 1997, Frail et al. 1997, 
Metzger et al. 1997) and detailed successful models (M\'{e}sz\'{a}ros \&
Rees 1997), bursts are now generally thought to be relativistically
expanding 
fireballs at cosmological distances.  The Lorentz factor of the expansion 
$\Gamma$ is generally thought to be from 100 to 1000 so as to explain 
the GeV photons seen in some bursts (Harding \& Baring 1994).  Within
this basic scheme, pulse durations and fall times can limit fireball 
properties (Fenimore, Madras, \& Nayakshin 1996;FMN) as can the rise
times. The model constraints will depend on the particular scenario
invoked, but 
some general arguments can use the rise times to constrain fireball 
properties independent of the specific scenario.

The first constraint is that the size of the central engine is limited to 
$c\tau_{min}$.  Within fireball 
models, the initial uncollimated flow will result in a density gradient at 
the front edge of the expanding shell with a width equal to the light travel 
time across the emission region.  For external shock scenarios, the 
fuzziness of the shell will result in emission starting to rise when 
the leading edge first hits the stationary cloud while the peak comes later 
 when the bulk of the shell hits the cloud.  For internal shock scenarios, 
the constraints will only be stricter since the outer shell is also moving. 
So the energy generation volume must have a typical thickness of smaller
than 1200 km for the majority of bursts.  A narrowly collimated jet
scenario might be able to substantially violate this limit.

The second constraint is on the physical dimension in the direction
perpendicular to the expansion of the shell. The arrival time for photons
from a single shell will be the travel time of the shell to the radius of
impact plus the travel time of the gamma ray to Earth. As the shell
expands at very close to light speed, the delay is purely geometrical,
with photons from off-axis regions being delayed compared to 
photons from on-axis regions.  The observed delay depends only on 
the radius of the shell at the time of impact with the cloud (R) and the 
angular radius of the gamma ray emission region as subtended from the burst 
site ($\Delta \Theta$). At a typical off axis angle such as $\Gamma^{-1}$,
the rise time will be close to $R(c\Gamma)^{-1}\Delta\Theta$. The
shell has been expanding for at least the time from the start of the
burst until the time of the rise ($T_{rise}$), so $R > 2c\Gamma^{2}
T_{rise}$ (FMN). Then, $\Delta \Theta< \tau_{rise}/(2\Gamma T_{rise})$.
For the bursts in our sample, $\tau_{rise} \sim 4 ms$, $T_{rise} > 0.1s$,
and $\Gamma > 100$, we find that $\Delta \Theta < 0.0002$ radians or $<
42$ arc-seconds. Due to self shadowing, Earth 
can only see a `cap' of the shell which subtends an angle $\Theta_{cap}
= \Gamma^{-1}$, so the individual emission region associated with the
rises substends only small region of the cap ($\sim 42$ arcseconds). This
is in contrast to the total fraction of the shell which becomes active,
$\sim 5 \times 10^{-3}$ (Fenimore et al. 1998). This demonstrates that
either the shell or the impacted cloud is very fragmented.

The third constraint is on the velocity dispersion within a single
individual emitting region. Based on the precedent of supernova shells, we 
expect there should be a substantial range of velocities within a shell,
with the fast moving material sorting itself to the front and the slow
moving ejecta in the rear.  For the external shock scenario, the flare
will start to brighten when the leading edge hits the cloud and will peak
when the bulk of the  shell hits the cloud, resulting in a measurable rise
time.  Let $\Gamma$ be the Lorentz factor for the densest layer of the
shell, with $\Delta \Gamma$ the difference in Lorentz factor between
densest layer and the leading edge of the shell. To 
account for the observed rise time, the fractional dispersion in Lorentz 
factors ($\Delta \Gamma / \Gamma$) within an emitting region must be less
than $\tau_{rise}/2T_{rise}$. For the majority of bursts, the $\Gamma$
dispersion is $\sim 2\%$ for the emitting regions. In contrast, the range
of $\Gamma$'s associated with the different emitting regions can have a
large dispersion, more than a factor of 2 (see Fenimore et al. 1998).

The fourth constraint is on the size scale of the impacted cloud
along the line of sight within  
the external shock scenario.  For a thin shell, the gamma radiation will start  
when the shell sweeps across the inner boundary of the cloud while the peak 
flux will be produced when the shell sweeps across the center (or densest 
region) of the cloud.  The characteristic dimension for the structure of the 
cloud must be smaller than $2\Gamma^{2}c\tau_{rise}$ since the shock is
going at near light speed (FMN).  For the average rise time of 4 ms and
$\Gamma < 1000$, the typical 
cloud size must be smaller than 16 AU.
  
The main conclusion from our research is that the majority of GRBs contain 
rises faster than four milliseconds in their light curves, and this places 
severe limits on burst scenarios. In particular, the size of the central 
engine region must be typically smaller than 1200 km. The individual gamma 
ray emitting region must be quite small (subtending only about 42
arc seconds). There can only be a small dispersion of $\Gamma$ factors
within the individual emitting regions.

\clearpage

\begin{table}
\begin{center}
\begin{tabular}{|c|c|c|c|c|c|}
\hline
GRB & Trigger & $P_{64}$ ($ph \cdot s^{-1}$) & $T_{90}$ (s) & TTE duration
(s) & $<C[32]>$ \\ 
\hline
910503 &143 &52 &50.8 &0.89 &0.98 \\
910609 &298 &56 &0.45 &1.10 &0.69 \\
910627 &451 &17 &15.2 &1.50 &0.55 \\
910718 &551 &5.6 &0.25 &0.94 &0.71\\
911109 &1025 &18 &2.62 &1.34 &0.59\\
911202 &1141 &9.3 &20.1 &1.50 &0.54 \\
920229 &1453 &12 &0.19 &1.42 &0.61\\
920622B &1664 &10.5 &3.52 &1.78 &0.46\\
920718 &1709 &14 &3.46 &0.85 &0.99 \\
920720 &1711 &22 &5.95 &0.73 &1.11 \\
921022 &1997 &40 &60.2 &0.78 &0.91 \\
930131 &2151 &168 &19.2 &0.078 &10.97 \\
930506 &2329 &43 &22.1 &1.49 &0.55 \\
930706 &2431 &44 &2.78 &0.54 &1.61 \\
930905 &2514 &28 &0.20 &0.66 &1.35 \\
930922 &2537 &27 &4.80 &0.56 &1.43 \\
931031 &2611 &35 &12.2 &0.65 &1.27 \\
950211 &3412 &55 &0.068 &1.17 &0.80 \\
950325B &3480 &22 &9.1 &0.43 &2.07 \\
950503 &3537 &... &$\sim10$ &0.40 &2.22 \\
\hline
\end{tabular}
\caption{Bursts analyzed.} 
\end{center}
\end{table}

\begin{table}
\begin{center}
\begin{tabular}{|c|c|c|c|}
\hline
GRB & $\tau_{min}$ (ms) & $T_{peak}$ & $A_{norm}$ (33ms) \\
\hline
910503 &2.0 &... &3.5 \\
910609 &1.0 &... &250 \\
910627 &33 &... &1.4 \\
910718 &1.0 &... &65 \\
911109 &16 &... &2.6 \\
911202 &33 &... &1.14 \\
920229 &0.26 &... &65 \\
920622B &4.1 &... &3.7 \\
920718 &33 &... &1.8 \\
920720 &4.1 &... &4.2 \\
921022 &0.51 &33 &477 \\
930131 &1.0 &... &... \\
930506 &8.2 &8.2 &1.17 \\
930706 &4.1 &66 &8.9 \\
930905 &1.0 &16 &20 \\
930922 &4.1 &... &34 \\
931031 &16 &66 &10.3 \\
950211 &4.1 &66 &330 \\
950325B &8.2 &... &9.4 \\
950503 &2.0 &33 &110 \\
\hline
\end{tabular}
\caption{Results from wavelet analysis.}
\end{center}
\end{table}

\begin{figure}
\begin{center}
\resizebox{12cm}{10cm}{\includegraphics{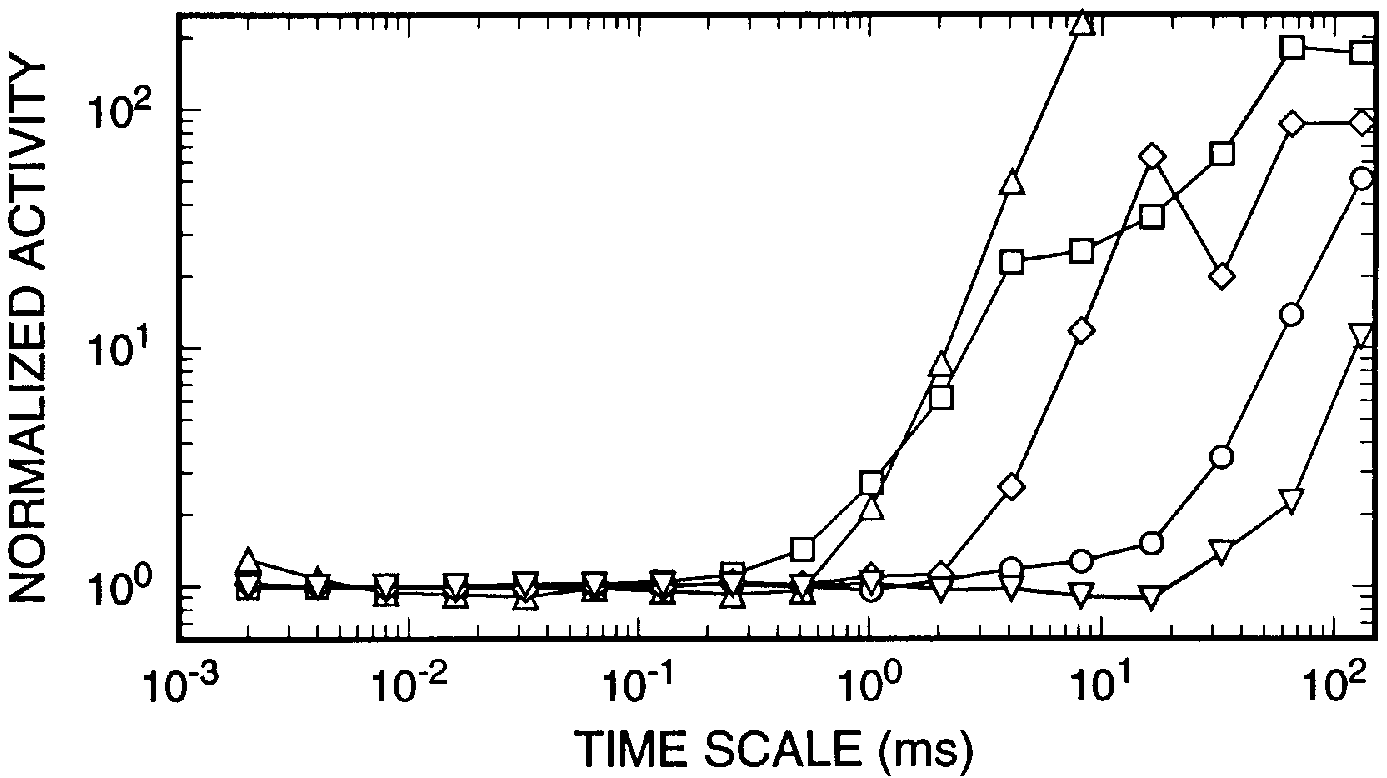}}
\caption{Normalized wavelet activity for five bursts. On each time
scale, the observed wavelet activity is divided by the expected activity
from normal Poisson fluctuations to get the normalized activity.  For each
of the five sample bursts, the normalized activity is close to unity for
time scales less than some $\tau_{min}$ value and then starts rising fast
for time scales longer then $\tau_{min}$. The $\tau_{min}$ values are when
the activity has risen 3-$\sigma$ above the Poisson level, and represents
primarily the rise times
in the light curves. The observed times of fastest variations range from
$256 \mu s$ to 33 ms, with the majority of bursts showing activity on the
four millisecond time scale. The upward triangles are for GRB930131,
squares for GRB 920229, diamonds for GRB930905, circles for GRB910503, and
downward pointing triangles for GRB910627. These results show that
millisecond variability is a common property of bursts, and thus provide
general constraints applicable to burst models.}
\end{center}
\end{figure}


\begin{thebibliography}{}

\bibitem{barat} Barat, C. et al. 1984, ApJ, 285, 791

\bibitem{bhat} Bhat, P. N. et al. 1992, Nature, 359, 217

\bibitem{cline} Cline, T. L. et al. 1980, ApJ, 237, L1

\bibitem{costa} Costa, E. et al. 1997, Nature, 387, 783

\bibitem{daubechie} Daubechies, I. 1992, Ten Lectures on Wavelets
(Philadelphia: Capital City Press)

\bibitem{deng} Deng, M. \& Schaefer, B. E. 1997, ApJ, 491, 720

\bibitem{fenimore96} Fenimore, E. E., Madras, C. D., \& Nayakshin, S.
1996, 473, 998 (FMN)

\bibitem{fenimore98} Fenimore, E. E., Cooper, C., Ramirez, E., Summer, M.
C., Yoshida, A., \& Namiki, M., 1998, ApJ, in press (astro-ph/9802200)

\bibitem{fishman} Fishman, G. J. et al. 1994, ApJSupp, 92, 229

\bibitem{frail} Frail, D. A. et al. 1997, Nature, 389, 261

\bibitem{giles} Giles, A. B. 1997, ApJ, 474, 464

\bibitem{harding} Harding, A. K. \& Baring, M. G. 1994, in Gamma-ray
Bursts, ed. G. Fishman et al. (New York; AIP 307), 520

\bibitem{kilaczyk} Kolaczyk, E. D. 1997, ApJ, 483, 34

\bibitem{kouveliotou} Kouveliotou, C. et al. 1994, ApJ, 422, L59

\bibitem{mazets} Mazets E. et al. 1981, Ap\&SS, 80, 3

\bibitem{metzger} Metzger, M. R. et al. 1997, Nature, 387, 476

\bibitem{meszaros} M\'{e}sz\'{a}ros, P. \& Rees, M. J. 1997, ApJ, 476, 232

\bibitem{mitrofanov} Mitrofanov, I. G. 1989, Ap\&SS, 155, 141

\bibitem{nemiroff} Nemiroff, R. J. et al. 1994, ApJ, 423, 432

\bibitem{norris} Norris, J. P. et al. 1994, ApJ, 424, 540

\bibitem{scargle} Scargle, J. 1997, in Applications of Time Series 
Analysis in Astronomy and Metrology, Chapman \& Hall, p. 226

\bibitem{schaefer01} Schaefer, B. E. et al. 1992, ApJ, 404, 673

\bibitem{schaefer02} Schaefer, B. E. \& Walker, K. C. 1998, ApJ, submitted

\bibitem{vanparadijs} van Paradijs, J. et al. 1997, Nature, 386, 686

\end{thebibliography}
\end{document}